\documentclass[11pt]{article}

\usepackage{amsmath,amssymb}
\usepackage{graphicx}
\usepackage{cite}
\usepackage{hyperref}
\usepackage{geometry}
\usepackage{authblk}
\usepackage{float}
\title{\bf The Dead Cone Effect in Heavy-Quark Jets:\\ A Unified Study from Charm and Bottom to Top
\footnote{Talk given by Redamy Perez-Ramos at the International Workshop on Future Linear Colliders (LCWS---2025), Valencia, Spain}
}

\author[1,2]{Redamy Perez-Ramos}
\author[3]{Stefan Kluth}
\author[3]{Wolfgang Ochs}

\affil[1]{DRII-IPSA, 63 Boulevard de Brandebourg, 94200 Ivry-sur-Seine, France}
\affil[2]{Laboratoire de Physique Th\'eorique et Hautes Energies (LPTHE), UMR 7589\\
Sorbonne Universit\'e et CNRS, 4 place Jussieu, 75252 Paris Cedex 05, France}
\affil[3]{Max-Planck-Institut f\"ur Physik, Boltzmannstr.\ 8, 85748 Garching, Germany}

\date{January 20, 2026 \hspace{1cm} MPP-2026-3} 

\begin{document}

\maketitle

\begin{abstract}
We present a unified overview of recent progress in the study of QCD radiation in heavy-quark jets, focusing on the dead-cone effect. Using precision data from LEP at $\sqrt{s}=91.2$~GeV, we demonstrate strong momentum-space suppression in charm and bottom quark jets, supported by Monte Carlo simulations with \textsc{Pythia}8, and provide a quantitative interpretation within the Modified Leading Logarithmic Approximation (MLLA) of perturbative QCD. We then extend the analysis to top-quark jets at $\sqrt{s}=1$~TeV, where finite lifetime effects and decay radiation introduce new conceptual challenges. A new method is presented to isolate the top-quark dead cone by separating production and decay radiation, and it is validated at both parton and hadron level using \textsc{Pythia}8. Together, these results establish a coherent framework for testing QCD radiation dynamics across all three heavy quarks. 
\end{abstract}
\newpage

\section{Introduction}


The suppression of gluon radiation at small angles from energetic heavy quarks, known as the \emph{dead cone effect}, is a fundamental prediction of perturbative Quantum Chromodynamics (QCD). For a heavy quark of mass $m_Q$ and energy $E_Q$, gluon emission is suppressed for angles $\Theta\lesssim\Theta_0$ with $\Theta_0=m_Q/E_Q$~\cite{Dokshitzer:1991fc,Dokshitzer:1991fd}. This effect modifies both the angular structure and the momentum spectra of particles produced by heavy-quark initiated jets. The angular distribution of gluons from the splitting process $Q\to Q+g$ is obtained as 
\begin{equation}
  d\sigma_{Q\to Q+g} \simeq \frac{\alpha_S}{\pi}C_F\frac{\Theta^2d\Theta^2}{(\Theta^2+\Theta^2_0)^2} \frac{d\omega}{\omega},
\label{emission}
\end{equation}
where $\alpha_S$ is the strong coupling constant and $C_F=4/3$ is the QCD colour factor~\cite{Dokshitzer:1991fc,Dokshitzer:1991fd}. With large emission angles $\Theta\gg \Theta_0$ the gluon radiation angular and momentum spectra are predicted to become identical to those of a light quark. 
By comparing the production of the heavy quarks c, b, and t, one can explore QCD radiation dynamics over more than two orders of magnitude in quark mass.
The first evidence for the dead cone effect has been reported by the ALICE collaboration at the LHC on c-quark jets based on an angular analysis and QCD MC simulations \cite{ALICE:2021aqk}. In our subsequent analysis the effect has been established in momentum distributions of hadrons for c- and b-quark jets in direct comparison with QCD-MLLA predictions~\cite{Kluth:2023umf}.

\section{Dead Cone Effect in Charm and Bottom Quark Jets at LEP}
The dead-cone effect in charm and bottom quark jets has been investigated using DELPHI and OPAL data at the $Z$ pole~\cite{DELPHI:1998cgx,OPAL:1998arz}. The inclusive momentum spectra of charged particles were corrected for contributions from heavy-hadron decays 
using \textsc{Pythia}8 \cite{Bierlich:2022pyt,Sjostrand:2014zea} simulations normalized to the measured decay multiplicities, yielding the genuine heavy-quark fragmentation functions. These distributions reveal a pronounced suppression of particle production at large momentum fraction in charm and 
bottom quark jets relative to light quark jets ~\cite{Kluth:2023umf,Kluth:2023dav,Kluth:2023rst}.

\subsection{Quantitative MLLA Interpretation}

While the MLLA framework provides a well-established description of hadron multiplicities in heavy quark events, a fully rigorous derivation of inclusive energy spectra in such events is still lacking. An approximate formulation has nevertheless been proposed, which reproduces the known MLLA results for multiplicities after integration over momentum fraction $x$ and ensures positivity of the fragmentation function. The corresponding MLLA-inspired expression for the fragmentation function reads~\cite{Dokshitzer:1991fd}
\begin{equation}
  \bar D_Q(\xi,W)=\bar D_q(\xi,W) -  \bar D_q( \xi - \xi_Q,\sqrt{e}M_Q),
\label{MLLAeqxi}
\end{equation}
where $W=\sqrt{s}$ is the $e^+e^-$ c.m.s.\ energy, $W_0 = \sqrt{e}\, M_Q$, $\xi=\ln(1/x)$ with $x=E_g/E_Q$ and $\langle x_Q \rangle$, the mean momentum fraction of the primary heavy quark such that, $\xi_Q = \ln(1/\langle x_Q \rangle)$. We take the experimental values $W_0=8$ GeV with
$\langle x_b \rangle\approx0.7$ for bottom jets and $W_0=2.7$ GeV with $\langle x_c \rangle\approx0.5$ for charm jets. Using experimental input for the light-quark spectra at energies $W$ and $W_0$, this relation reproduces the observed suppression patterns for both charm and bottom jets in the central kinematic region. At very large $\xi_p$, an excess of particle production is observed in bottom jets relative to MLLA expectations, indicating the onset of non-perturbative dynamics in the ultrasoft region.
\begin{figure}[t]
  \centering
  \includegraphics[width=0.48\textwidth]{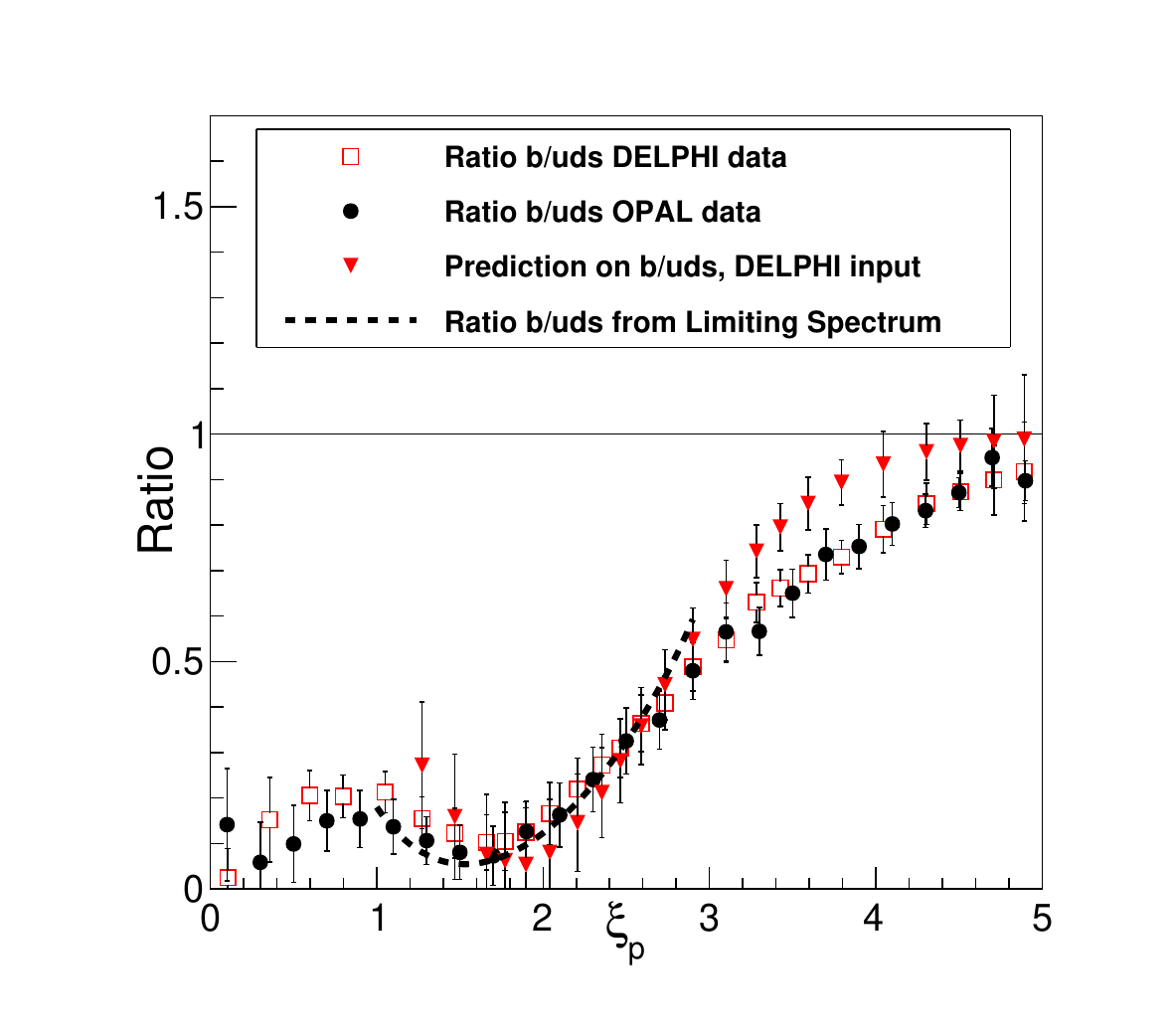}\hfill
  \includegraphics[width=0.515\textwidth]{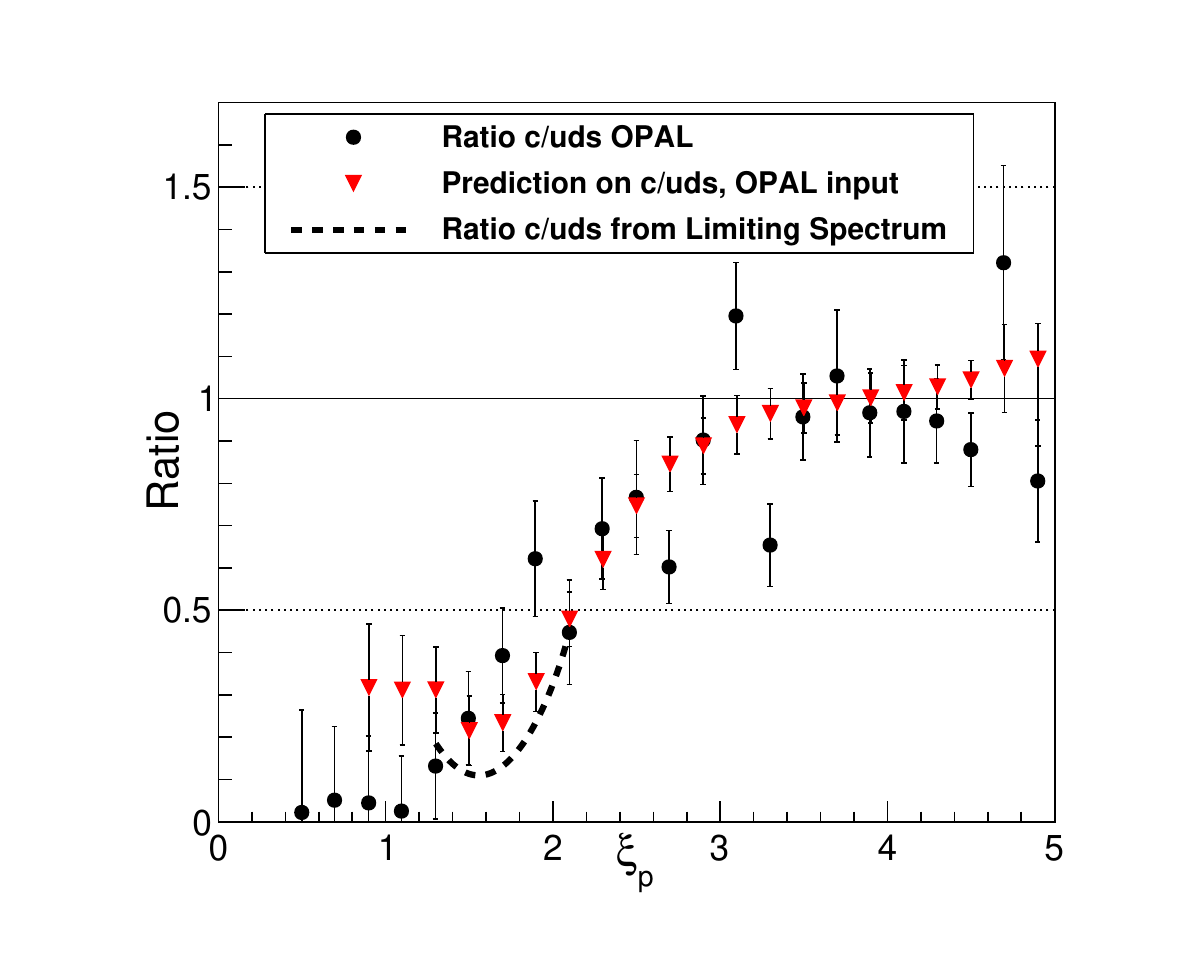}
  \caption{Ratio of the heavy $b$-quark to the light $uds$-quark fragmentation functions (left) and the corresponding ratio for the $c$ quark (right), compared with the MLLA expectations in Eq. \eqref{MLLAeqxi} based on experimental data, Limiting Spectrum distributions, and \textsc{Pythia}~8.}
  \label{fig:exp-deadcone_ratios}
\end{figure}
In Fig.~\ref{fig:exp-deadcone_ratios} we display the ratio of heavy quark over light quark fragmentation functions in comparison with the MLLA prediction eq.~\eqref{MLLAeqxi} using either data as input or fits to the ``Limiting Spectrum", an analytical representation of momentum spectra within MLLA \cite{Dokshitzer1991Basics}. These ratios are correctly reproduced for b-quarks in the central region $1\lesssim\xi_p\lesssim 3$. For c-quarks the predictions from the data are generally compatible with the measurements within the large errors whereas the predictions from the Limiting Spectrum agree with the data within the accessible region $1\lesssim\xi_p\lesssim 2$. 

\section{Identifying the Dead Cone in Top-Quark Jets}

\subsection{Conceptual Challenges}

Extending the study of the dead cone to top-quark jets introduces qualitatively new challenges which have been studied in Ref.~\cite{Kluth:2025xpj} using the \textsc{Pythia}~8 MCEG~\cite{Bierlich:2022pyt,Sjostrand:2014zea}. The large top mass shifts the dead-cone angle to considerably larger values, but the finite lifetime of the top quark and the radiation from its decay products obscure the direct observation of the primary radiation pattern. This approach could be tested beyond the production threshold of the top anti-top pair production in the Future Circular Collider (FCC), but with additional modifications also tests at the pp collider are possible. In this framework, we consider the top quark decay according to the Feynman diagrams displayed in Fig.~1 of Ref.~\cite{Kluth:2025xpj} for the splittings $t\to b\ell\nu$ and $t\to b\ell\nu+g$. The b-quark radiates gluons as well, and the observed radiation pattern becomes a superposition of production and decay contributions, which is given in $O(\alpha_s)$ as 
~\cite{Orr:1992uv}
\begin{gather} 
\frac{1}{\sigma_0} \frac{d\sigma_g}{d\omega\ d\cos\theta_g d\phi_g} = \frac{\alpha_s C_F}{4\pi^2} \omega F, \label{FKhozeOtt} \\
F=\lvert A\rvert^2 + \  \lvert B_1\rvert^2 \ + \   \lvert B_2\rvert^2 \ - 2Re[B_1B_2^*] \ +  2Re[A(B_1-B_2)^*]. \nonumber
\end{gather}
The amplitudes $A$, $B_1$ and $B_2$ are constructed from the external momenta of the quarks $p_i$ and the emitted gluon $k$ and can be associated with the colour dipoles $\widehat{t\bar t}$, $\widehat{tb}$ and $\widehat{\bar t \bar b}$, respectively~\cite{Dokshitzer1991Basics}. The contribution proportional to $|A|^2$ describes gluon radiation from the top quark in the production stage, as in Eq.~\eqref{emission}. The term $|B_1|^2$ accounts for gluon emission accompanying the decay process $t \to bW$, where radiation can originate from either the $t$ or $b$ quark, while the analogous contribution $|B_2|^2$ describes the corresponding radiation in the $\bar t$ decay channel. This contribution approximately vanishes for b-quarks emitted in top-quark direction. The amplitudes are rewritten in terms of the variables $X = (\Theta/\Theta_0)\cos\phi$ and $Y = (\Theta/\Theta_0)\sin\phi$, where $\Theta$ is the angle between the considered parton and the top-quark direction, and $\phi$ is the corresponding azimuthal angle around the top quark flight direction and the azimuth is chosen to take the value $\phi=\pi$ for the direction of the b-quark. In the following, $X_B$ is defined as the angle between the $b$-quark jet and the original top-quark direction prior to its decay (see next section).

Another key step in the analysis is the reconstruction of the invariant masses of the top quark from final-state hadrons, $M(B\ell\nu)$ and $M(B\ell\nu+g)$, corresponding to the splittings $t \to b\ell\nu$ and $t \to b\ell\nu+g$, respectively. This procedure results in two narrow bands in the plane $(M(B\ell\nu),M(B\ell\nu+g))$ around the top-quark mass $m_t \simeq 172.5$~GeV~\cite{Kluth:2025xpj,MaltoniSelvaggiThaler:2016DeadCone}, with a width governed by the top-quark decay width $\Gamma_t = 1.42^{+0.19}_{-0.15}$~GeV~\cite{ParticleDataGroup:2024cfk}.

\subsection{Separation of Production and Decay Radiation: Extrapolation to $\boldsymbol{X_B=0}$}

\begin{figure}[t]
  \centering
  \begin{minipage}{0.60\linewidth}
    \centering
    \includegraphics[width=\linewidth]{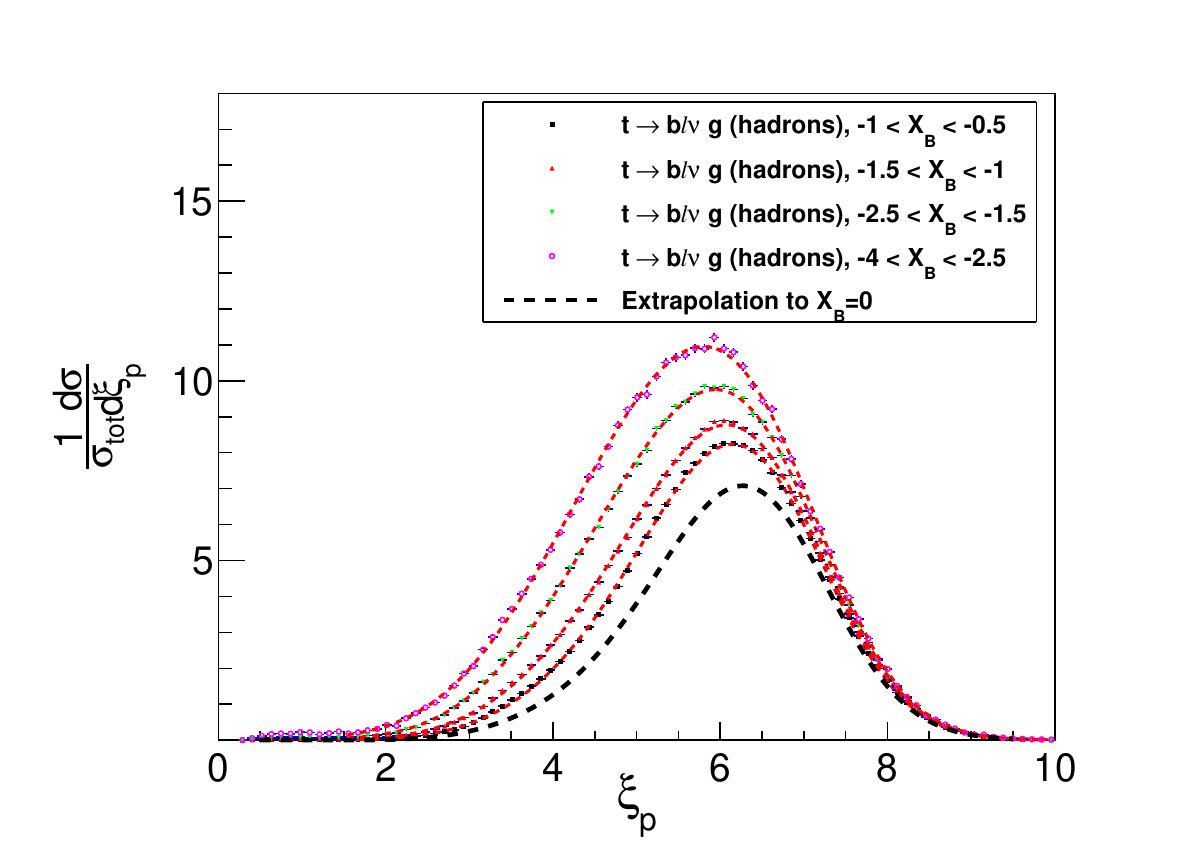}
  \end{minipage}\hspace{0.02\linewidth}
  \begin{minipage}{0.30\linewidth}
    \centering
    \includegraphics[width=\linewidth]{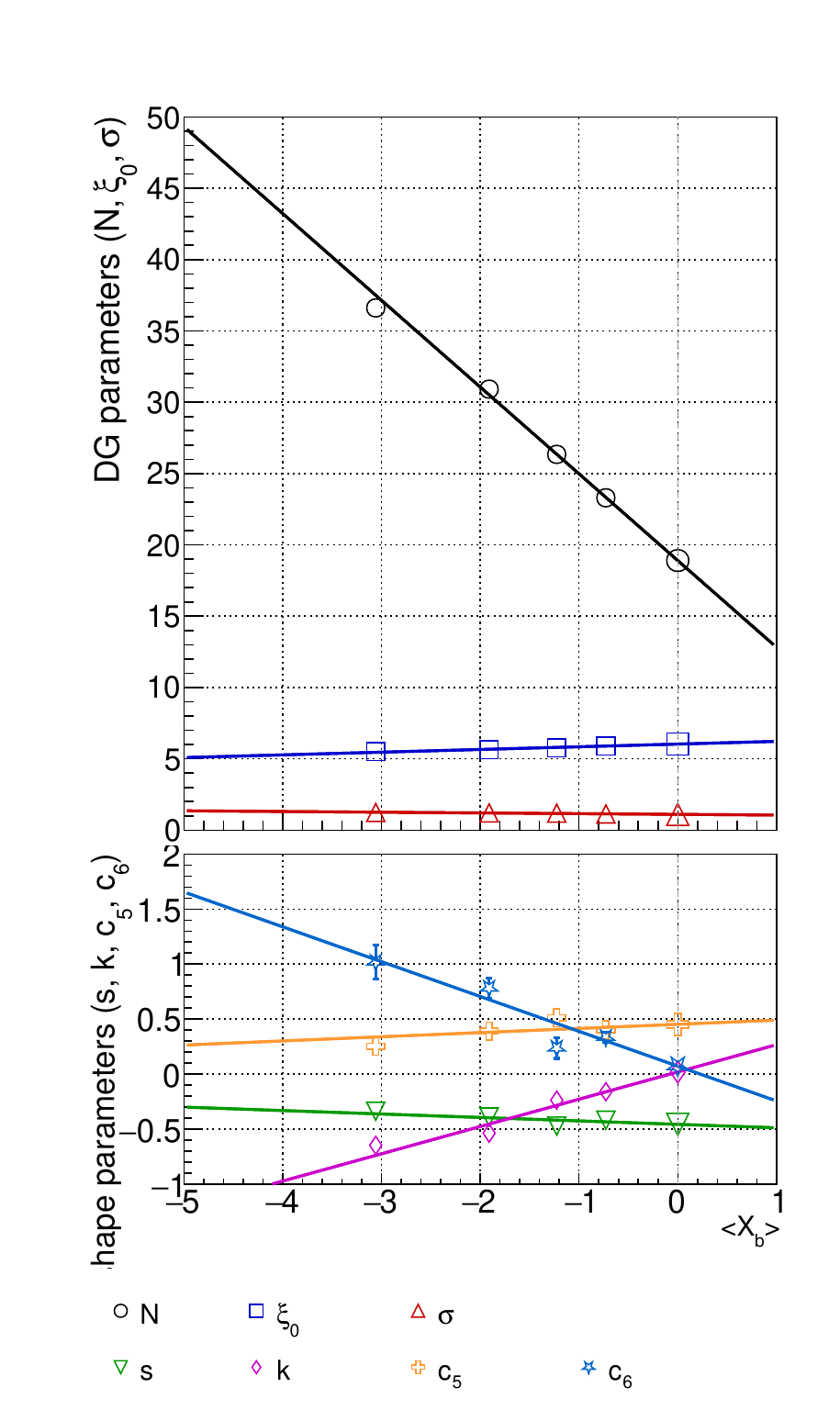}
  \end{minipage}
  \caption{Distribution of $\xi=\ln(1/x_p)$. Left: $\xi$ distributions for different intervals of the decay angle $X_B$ of the $b$ quark from top decay; the dashed line shows the extrapolation to $X_B=0$ with the $\widehat{tb}$ dipole radiation removed. Right: moment parameters of the Distorted Gaussian distribution and linear fits to the $X_B$ dependence, extrapolated to $X_B=0$.}
  \label{fig:multiple_xi_hadrons}
\end{figure}

A method has been developed to disentangle radiation from the production dipole and from the decay process. Guided by the leading-order QCD amplitudes and angular ordering, the strategy consists of selecting events with lateral $b$-quark emission into the left hemisphere ($X<0$) and extrapolating momentum distributions towards the forward direction $X_B=0$ , where the decay radiation vanishes. Events are selected within four angular regions, as shown in Fig.~\ref{fig:multiple_xi_hadrons} (left panel), and within the mass window $170 < M(B\ell\nu) < 175$~GeV. If the angle $X_B$ between the $b$-quark jet and the original top-quark direction increases, the $\xi$ distribution rises due to the growing contribution of the term $|B_1|^2$ associated with the $\widehat{tb}$ dipole in \eqref{FKhozeOtt}, its contribution is
removed by extrapolation to $X_B = 0$.
This extrapolation is performed using fits to the first seven moments of a distorted Gaussian distribution (multiplicity $N$, peak position $\xi_0$, width $\sigma$, skewness $s$, kurtosis $k$, and higher order moments $c_5$ and $c_6$), as shown in Fig.~\ref{fig:multiple_xi_hadrons} (right panel). The procedure enables the reconstruction of the radiation pattern corresponding to a hypothetical stable top quark, thereby isolating the dead-cone effect by removing the contribution from the $\widehat{tb}$ dipole effectively.

\subsection{Monte Carlo Validation and Momentum Suppression}

\begin{figure}[t]
\begin{center}
\includegraphics[height=5.5cm,width=7.0cm]{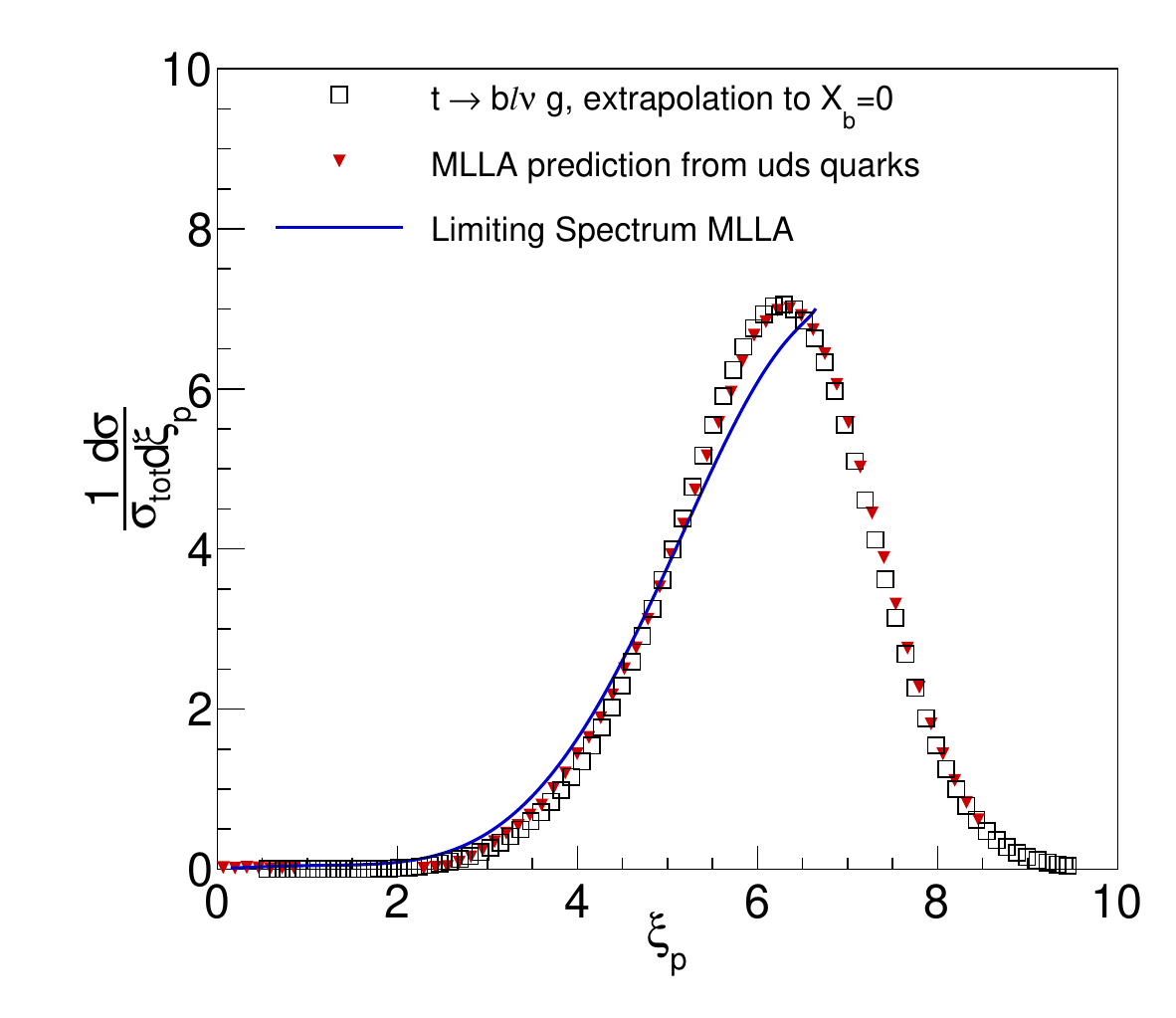}
\includegraphics[height=5.5cm,width=7.0cm]{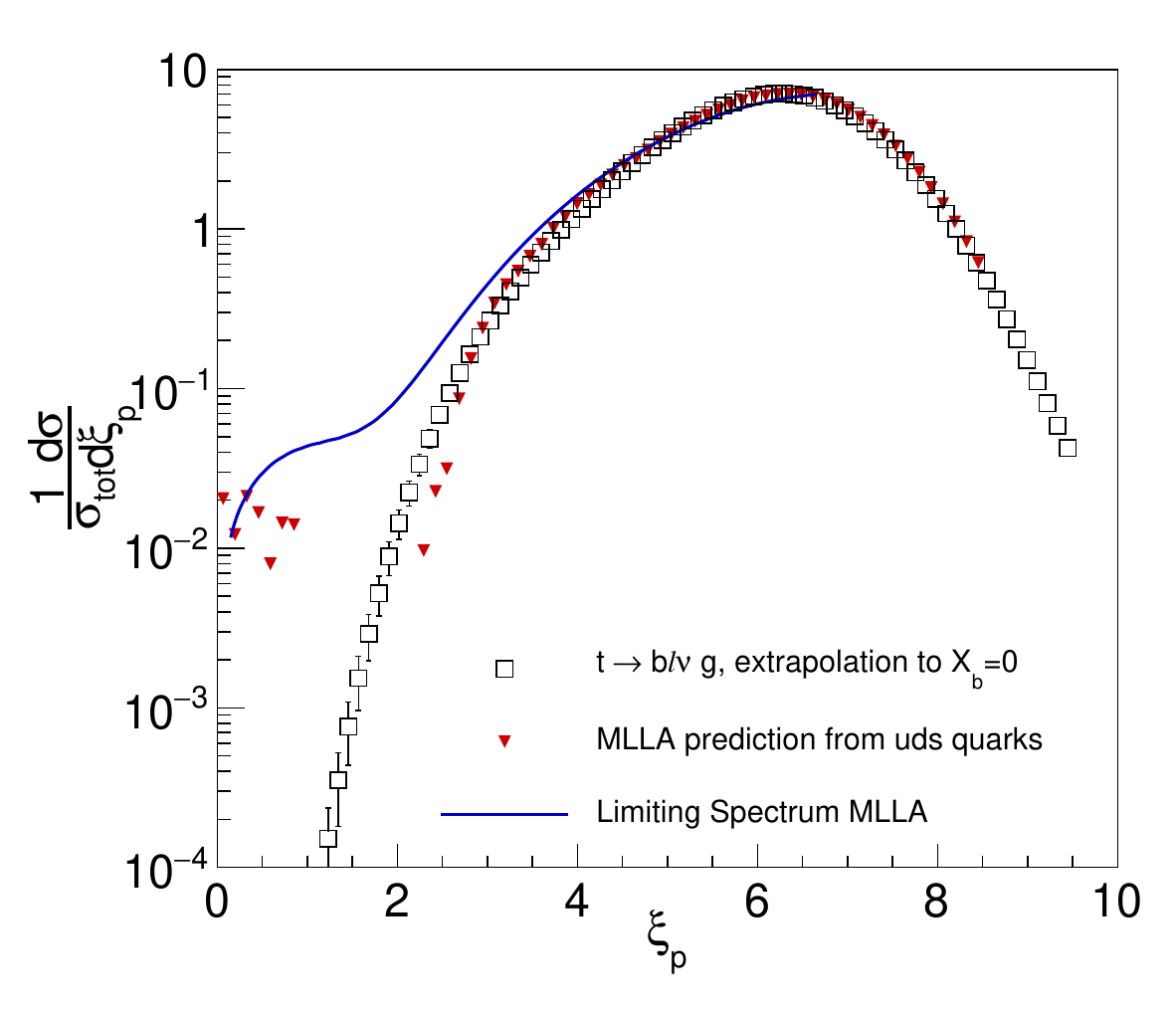}
  \caption{$\xi_p$ distributions for top-quark jets extrapolated to $X_B=0$ compared with MLLA expectations (left) and the corresponding distributions on a logarithmic scale (right).}
\label{fig:xi-dist-MLLA 1000}
\end{center}
\end{figure}

The method has been validated using \textsc{Pythia}~8.3 simulations at both parton and hadron level. In Fig.~\ref{fig:xi-dist-MLLA 1000}, the $\xi_p$ distribution of hadrons obtained by extrapolation to $X_B=0$ are compared with the MLLA predictions derived from light-quark spectra according to Eq.~\eqref{MLLAeqxi}, as well as with the approximate Limiting Spectrum predictions. 
Good agreement is observed over the central kinematic region, both in linear and logarithmic representations. In addition, the momentum spectra display a suppression of particles at small
$\xi$. For $\xi\gtrsim 2$ the extrapolated $\xi$ distributions are compatible with the MLLA expectations within an accuracy of about 10--15\%, providing an independent and experimentally robust signature of the dead-cone effect in top-quark jets.

\section{Unified Picture of Heavy-Quark Dead Cones}

\begin{figure}[t]
\begin{center}
\includegraphics[height=5.5cm,width=7.0cm]{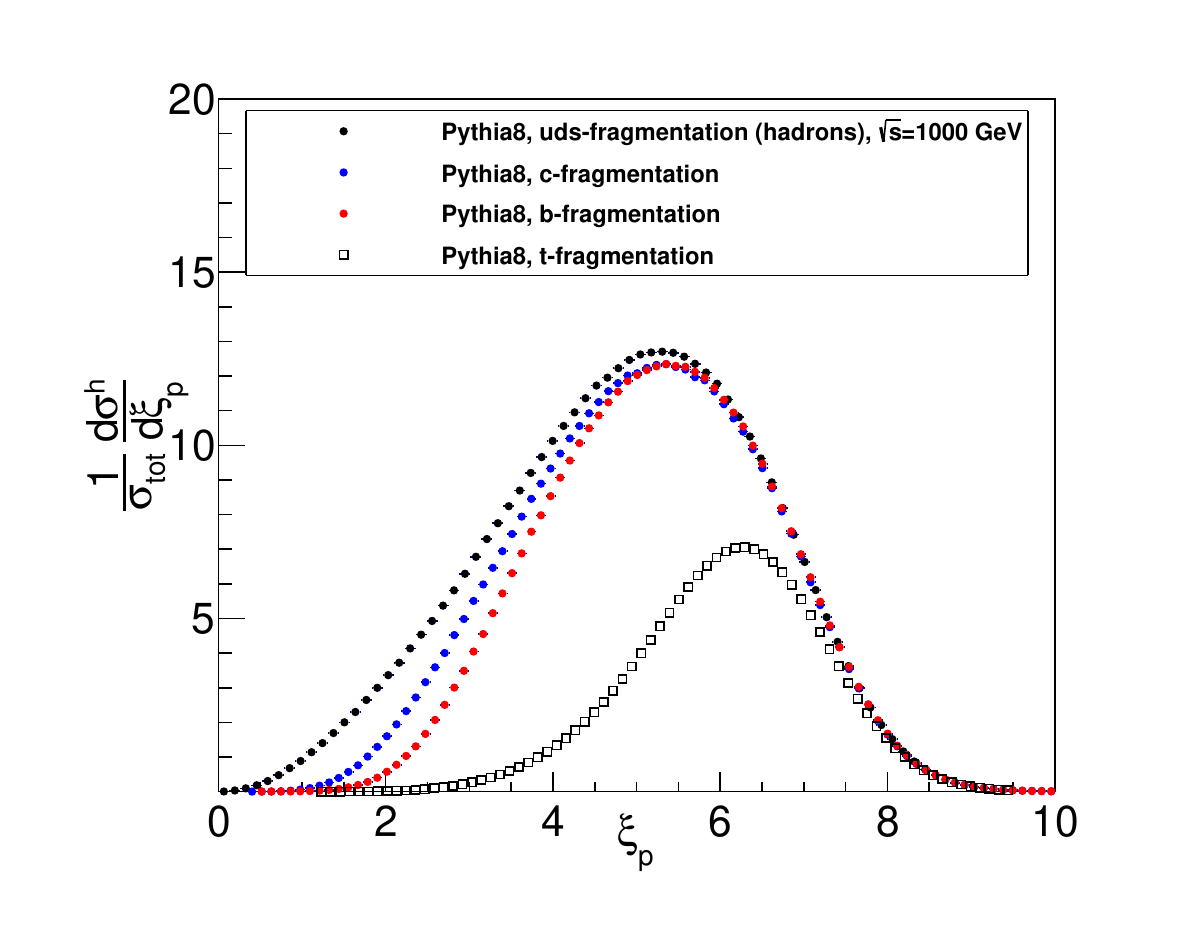}
\includegraphics[height=5.5cm,width=7.0cm]{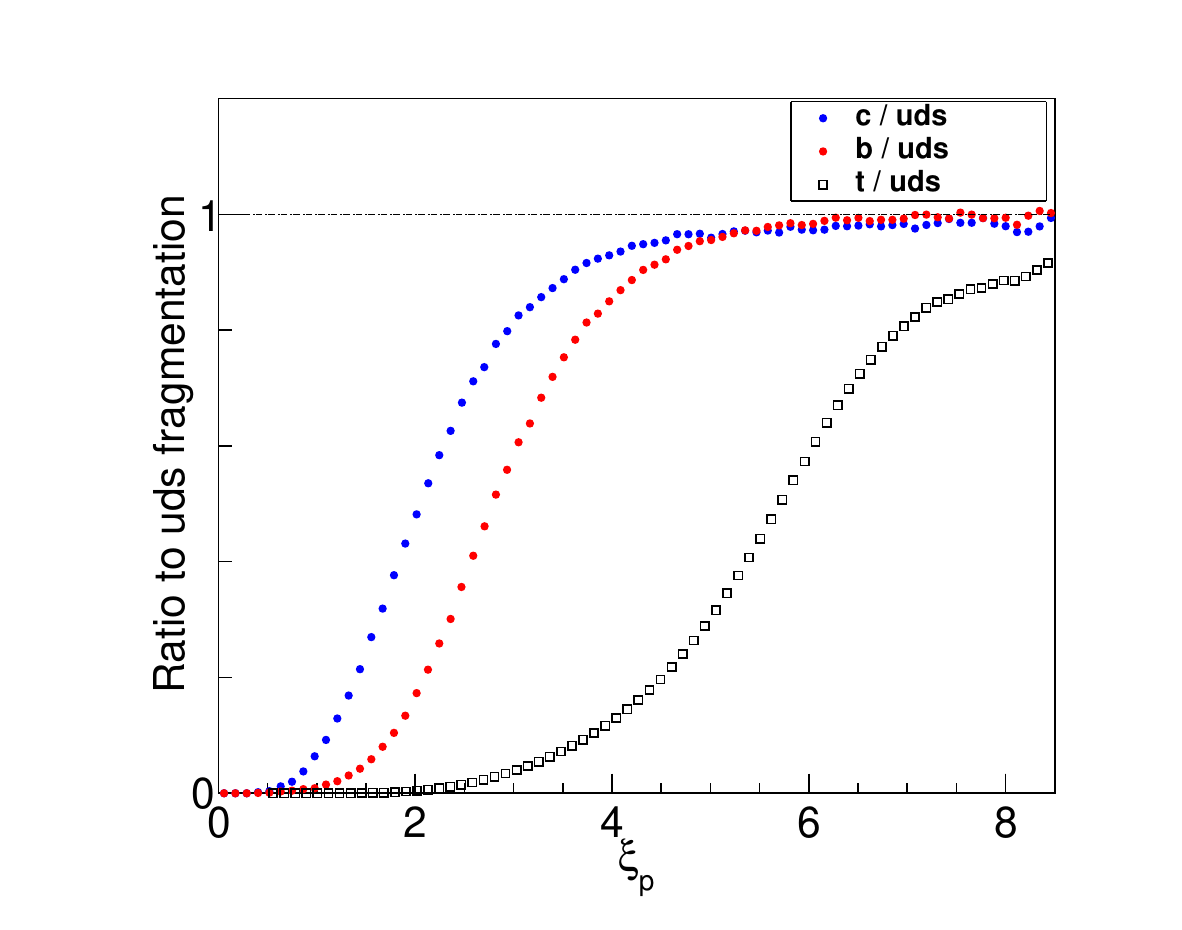}
  \caption{Comparison of the dead-cone suppression factor for charm, bottom, and top quarks at $\sqrt{s}=1000$~GeV. The hierarchical pattern illustrates the strong mass dependence of QCD radiation and the macroscopically large dead cone for the top quark.}
\label{fig:deadcone_cbt}
\end{center}
\end{figure} 

The combined studies of charm, bottom, and top quark jets provide a coherent and quantitative picture of mass effects in QCD radiation over more than two orders of magnitude in quark mass. 
At a centre-of-mass energy of $\sqrt{s} = 1000$~GeV, corresponding to a heavy-quark energy $E_Q \simeq 500$~GeV, the characteristic dead-cone angles follow the strong hierarchy
$\Theta_0^c \ll \Theta_0^b \ll \Theta_0^t$,
reflecting directly the mass ordering $m_c \ll m_b \ll m_t$. This hierarchy is clearly visible in the comparison of the $\xi$ distributions and in the corresponding ratios to light-quark fragmentation shown in Fig.~\ref{fig:deadcone_cbt}. The charm and bottom distributions remain relatively close to each other across the full $\xi$ range, both in the absolute spectra and in their ratios to $uds$ jets. This indicates that their dead cones are restricted to very small angles and primarily affect the hard part of the fragmentation.
In contrast, the top-quark fragmentation exhibits a dramatically stronger suppression, both in the $\xi$ distributions and in the ratios to light quarks. 
As a result of the large top-quark mass, the top-quark curve is clearly separated from the charm and bottom curves, demonstrating that the dead-cone effect is not a small correction but a dominant feature of top-quark radiation dynamics.
The smooth transition from the mild suppression in charm jets, through the stronger but still limited effect in bottom jets, to the pronounced depletion in top jets provides a striking theoretical illustration of the mass dependence predicted by perturbative QCD. In this sense, the dead cone emerges as a universal organising principle for heavy-quark radiation.

\section{Outlook}

The unified study of the dead cone across charm, bottom, and top quarks demonstrates the predictive power of perturbative QCD in describing radiation dynamics over a wide range of masses and lifetimes. The momentum-space suppression observed in charm and bottom jets and the newly developed method for top-quark jets together establish the dead cone as a robust and experimentally accessible QCD signature. 

The clear separation between the charm and bottom curves on the one hand and the strongly suppressed top quark distributions on the other hand highlights the unique role of the top quark as a laboratory for studying QCD radiation in the presence of a very large mass and a finite lifetime. In this respect, top-quark jets provide direct access to the interplay between production radiation, decay radiation, and colour coherence effects, opening new possibilities for precision tests of perturbative QCD.

Future measurements at high-energy lepton colliders, such as the FCC, and at the LHC, combined with modern jet substructure techniques, will allow these effects to be probed with increasing precision. In particular, differential studies of $\xi$ distributions, transverse-momentum spectra, and angular correlations in heavy-quark jets will make it possible to test the dead-cone dynamics beyond leading-logarithmic accuracy and constrain nonperturbative effects in the ultrasoft region.

From a broader perspective, the dead cone provides a unifying link between heavy-quark fragmentation, jet structure, and fundamental QCD radiation patterns.  In this way, the dead cone not only deepens our understanding of QCD dynamics but also contributes directly to the precision program of current and future collider experiments.

\newpage
\bibliographystyle{unsrt}

\bibliography{biblioproceedings}

\end{document}